\documentclass[a4paper]{jpconf}
\usepackage{graphicx}
\usepackage{amsmath}
\usepackage{color}
\usepackage{subfigure}
\usepackage{braket}
\usepackage{amssymb}

\begin{document}
\title{The next generation of laser spectroscopy experiments using light muonic atoms}


\author{S.~Schmidt$^{1}$, M.~Willig$^{1}$, J.~Haack$^{1}$, R.~Horn$^{1}$, A.~Adamczak$^{2}$, M.~Abdou Ahmed$^{3}$, F.~D.~Amaro$^{4}$, P.~Amaro$^{5}$, F.~Biraben$^{6}$,  P.~Carvalho$^{5}$, T.-L. Chen$^{7}$, L.~M.~P.~Fernandes$^{4}$, T.~Graf$^{3}$, M.~Guerra$^{4}$, T.~W.~H{\"a}nsch$^{8}$, M.~Hildebrandt$^{9}$, Y.-C.~Huang$^{7}$, P.~Indelicato$^{6}$, L.~Julien$^{6}$, K.~Kirch$^{9,10}$, A.~Knecht$^{9}$, F.~Kottmann$^{10}$, J.~J.~Krauth$^{1,8}$, Y.-W.~Liu$^{7}$, J.~Machado$^{5}$, M.~Marszalek$^{9}$, C.~M.~B.~Monteiro$^{4}$, F.~Nez$^{6}$, J.~Nuber$^{9}$, D.~N.~Patel$^{7}$, E.~Rapisarda$^{9}$, J.~M.~F.~dos Santos$^{4}$, J.~P.~Santos$^{5}$, P.~A.~O.~C.~Silva$^{4}$, L.~Sinkunaite$^{9}$, J.-T.~Shy$^{7}$, K.~Schuhmann$^{10}$, I. Schulthess$^{10}$, D.~Taqqu$^{10}$, J.~F.~C.~A.~Veloso$^{11}$, L.-B.~Wang$^{7}$, M.~Zeyen$^{10}$, A.~Antognini$^{9,10}$ and R.~Pohl$^{1,8}$}


\address{${^1}$Institut f\"ur Physik $\&$ Exzellenzcluster PRISMA, Johannes Gutenberg-Universit\"at Mainz, Germany. ${^2}$ Institute of Nuclear Physics, Polish Academy of Sciences, Poland. ${^3}$Institut f\"ur Strahlwerkzeuge, Universit\"at Stuttgart, Germany. $^{4}$LIBPhys, Department of Physics, University of Coimbra, Portugal. $^{5}$Laborat\'{o}rio de Instrumenta\c{c}\~{a}o, Engenharia Biom\'{e}dica e F\'{i}sica da Radia\c{c}\~{a}o (LIBPhys-UNL), Departamento de F\'{i}sica, Faculdade de Ci\^{e}ncias e Tecnologia, FCT, Universidade Nova de Lisboa, Portugal. $^{6}$Laboratoire Kastler Brossel, Sorbonne Universit\'{e}, CNRS, ENS-Universit\'{e} PSL, Coll\`{e}ge de France, Paris, France. $^{7}$Physics Department, National Tsing Hua University, Hsinchu, Taiwan. $^{8}$Max-Planck-Institut f\"ur Quantenoptik, Garching, Germany. $^{9}$Paul Scherrer Institute, Villigen, Switzerland. $^{10}$Institute for Particle Physics and Astrophysics, ETH Zurich, Switzerland. $^{11}$I3N, Departamento de F\'{i}sica, Universidade de Aveiro, Portugal.}

\ead{stefanschmidt@uni-mainz.de}

\begin{abstract}
Precision spectroscopy of light muonic atoms provides unique information about the atomic and nuclear structure of these systems and thus represents a way to access fundamental interactions, properties and constants. One application comprises the determination of absolute nuclear charge radii with unprecedented accuracy from measurements of the 2S\,-\,2P Lamb shift. Here, we review recent results of nuclear charge radii extracted from muonic hydrogen and helium spectroscopy and present experiment proposals to access light muonic atoms with $Z \geq 3$. In addition, our approaches towards a precise measurement of the Zemach radii in muonic hydrogen ($\mu$p) and helium ($\mu$$^{3}$He$^{+}$) are discussed. These results will provide new tests of bound-state quantum-electrodynamics in hydrogen-like systems and can be used as benchmarks for nuclear structure theories.
\end{abstract}

\section{Introduction}

In 1911, E.\,Rutherford resolved the internal structure of the atom for the first time \cite{Rut1911}, initiating nuclear physics. 
 Since then, the investigation of nuclear properties of the (lightest) elements has been crucial for our understanding of fundamental processes in nature. For example, it turned out that nuclear masses can provide detailed information about the stellar nucleosynthesis or neutrino physics \cite{Bla2006, Bla2013, Fil2016}. In parallel, the size of the atomic nuclei, i.e.~the root mean square (rms) charge radii, has established itself as a key parameter for tests of nuclear structure calculations \cite{Bla2013}, QED calculations \cite{Pac2003, Kar2005} and for the extraction of fundamental constants such as the Rydberg constant \cite{Moh2016, Poh2017, Bey2017, Fle2018}. Furthermore, it plays a crucial role for the explanation of the composition of nucleons inside a nucleus, leading to a number of cluster models (see e.\,g.~\cite{Nef2008}) or to the concept of halo nuclei \cite{Arn1992, Neu2008}.

At present, three complementary techniques are applied to obtain nuclear charge radii: elastic electron scattering \cite{Han1963, Ber2010, Zha2011, Sick2015}, high-precision laser spectroscopy of isotope shifts in regular atoms \cite{Bla2013} and X-ray spectroscopy of muonic atoms \cite{Wu1969, Ang2013}. Traditionally, elastic electron scattering has been the method of choice to determine the internal structure of nuclei. Hereby, the elastic scattering on the target nucleus is described by form factors included in the theoretical expression of the scattering cross section, which provide detailed information about the electric charge and the magnetic distributions inside the nucleus. Hence, they are used to determine absolute rms nuclear charge $r_{\rm{E}}$ and magnetic $r_{\rm{M}}$ radii, typically with uncertainties of a percent or slightly better \cite{Sick2015, Jan1972, Noe2011,  Ang2013}. Nowadays, scattering experiments using muons as a projectile are under construction \cite{Abb2007, Gil2013}. By a simultaneous determination of electron and muon scattering form factors, those experiments will allow a precise test of the lepton universality and thus will contribute to the so-called \textit{proton radius puzzle} \cite{Poh2010,Poh2013, Bern2014} in the near future.

Additionally, in the last two decades, laser spectroscopy of isotope shifts in regular atoms has proven to be a powerful tool for the determination of nuclear charge radius \textit{differences} \cite{Bla2013, Fri2003}. In addition to the most common technique of collinear laser spectroscopy \cite{Mue1983}, sophisticated methods such as two-photon or trap-assisted laser spectroscopy paved the way for the determination of charge radii of various elements including rare, short-lived isotopes \cite{Wan2004, San2006, Mue2007}. However, except for hydrogen-like atoms \cite{Moh2016, Bey2013}, theory is not yet precise enough to extract absolute charge radii from laser spectroscopy of atoms. Thus in contrast to electron scattering, only the difference of squared charge radii can be determined from isotope shift measurements, and one anchor nucleus with known size is required to calculate absolute charge radii from the differences.

X-ray spectroscopy of muonic atoms has for a long time been used to determine absolute charge radii for nuclei above carbon \cite{Ang2013}. In addition to that, in 1975 a direct measurement of the 2S\,-2P Lamb shift in muonic atoms via laser spectroscopy was proposed to be ideally suited for the determination of the size of the lightest nuclei ($Z\leq 5$) \cite{Zav1975}. The simple structure of these systems, a single muon orbiting a bare nucleus, allows for the determination of absolute charge radii with unprecedented accuracy. Having 207-times the electron mass, the muon spends more time inside the nucleus (enhanced overlap between the muon and the nucleus wave function) and thus acts as a sensitive probe for nuclear effects. 

\begin{figure}[h]
\centering
\includegraphics[width=0.95\textwidth]{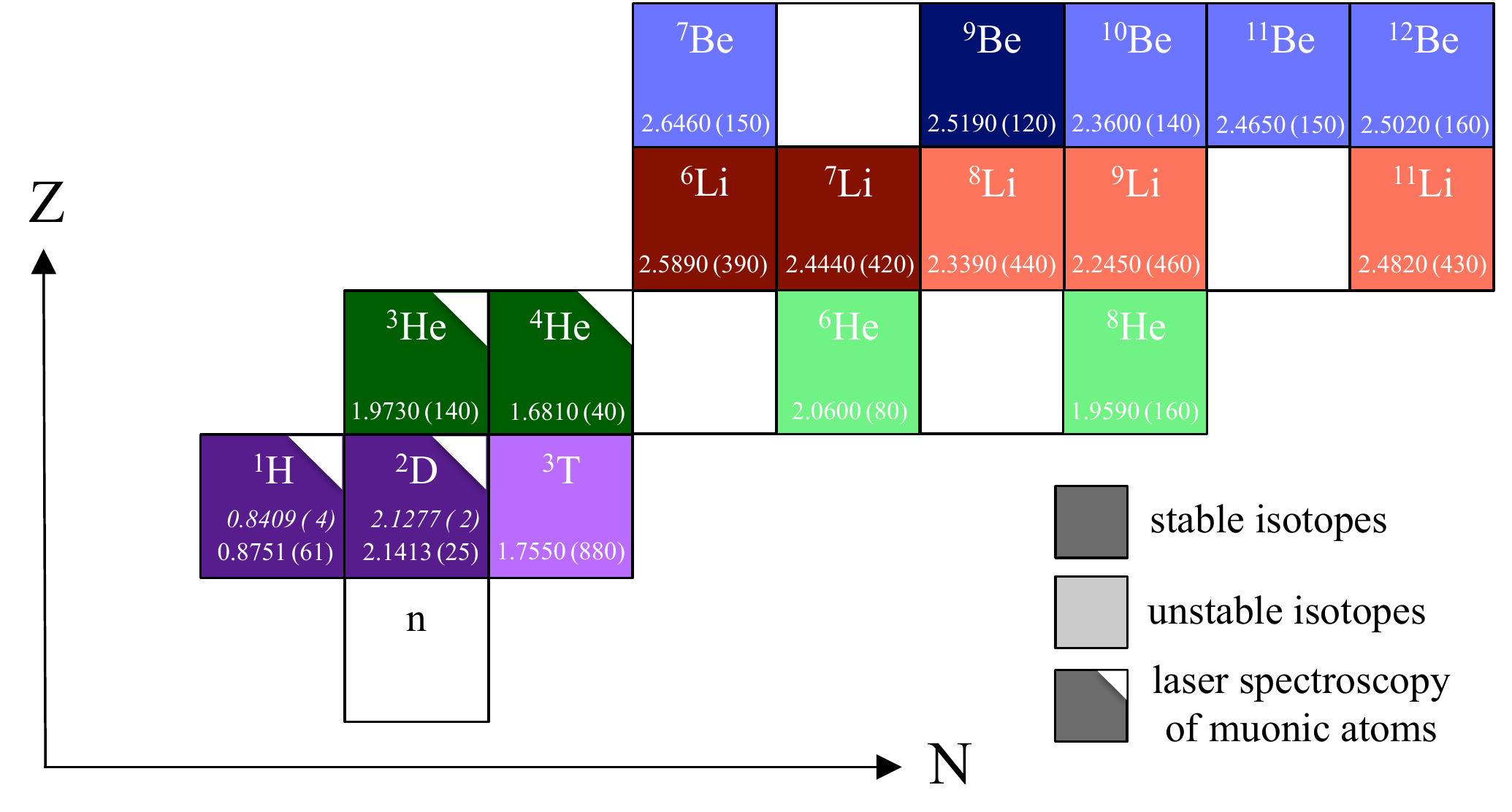}
\caption{\label{Fig0} A section of the table of nuclides including current literature values for the nuclear charge radii of the individual isotopes. The values printed in italic are extracted from muonic atom spectroscopy, whereas all other values result from either electron scattering or laser spectroscopy on regular atoms or both (CODATA-2014 values \cite{Moh2016}). All values are given in fm.}
\end{figure}

As summarized in Fig.\,\ref{Fig0}, laser spectroscopy of the Lamb shift in light muonic atoms has been used to extract absolute charge radii of the proton ($\mu$p) \cite{Poh2010} and the deuteron ($\mu$d) \cite{Poh2017} (values printed in italic). All other values presented in Fig.\,\ref{Fig0} result from either electron scattering or laser spectroscopy using regular atoms. As can be seen for hydrogen, the muonic values deviate by more than 5$\,\sigma$ from the CODATA-2014 value \cite{Ant2013}, which includes data from hydrogen spectroscopy and electron scattering. This discrepancy is known as the \textit{proton radius puzzle}. The result has triggered many new activities in the field, which are partially addressed in Sec.\,4\,-\,5.  Results from the measurement of the Lamb shift in the two stable helium isotopes ($\mu^{3}$He$^{+}$ and $\mu^{4}$He$^{+}$) will be published soon.

In this report we briefly review the results of nuclear charge radii of the lightest elements and present novel approaches to extend the laser spectroscopy experiments of muonic systems to the elements of lithium and beryllium. In addition, we focus on the determination of the Zemach radius of the proton and helion by precision spectroscopy of the ground-state hyperfine splitting in muonic hydrogen ($\mu$p) and muonic helium ($\mu^{3}$He$^{+}$) and discuss the current stage of future experiments.

\section{Hydrogen}


\subsection{Nuclear charge radii}

In atomic physics convention, the rms nuclear charge radius $r_{\rm{E}}$ of the \textbf{proton} is defined as
\begin{align}
\label{Eq1}
 \braket{r_{\rm{E}}^{2}}=-6\frac{d G_{\rm{E}}}{d Q^2}\Big{|}_{Q^2=0}\:\:\:\:\: \left(\sim \int\rho_{\rm{E}}(\textit{r}) \textit{r$^2$}  \rm{d}^3\textit{r}\right),
\end{align}
where $G_{\rm{E}}$ is the Sachs electric form factor of the nucleus, $Q^2$ is the negative of the square of the four-momentum transfer and $\rho_{\rm{E}}(\textit{r})$ is the charge density distribution of the nucleus. 
Historically, the Sachs electric form factor in Eq.\,\ref{Eq1} has been introduced as an extension of the photon-nucleus vertex in the Dirac theory to account for the finite size of the nucleus (see e.g.~\cite{Eid2007}). Starting from the modified interaction potential~\cite{Eid2007}
\begin{align}
\delta V(\textit{r})=V_{C}(\textit{r})-V_{C}^{pt}(\textit{r})=-4\pi\alpha\int\frac{d^3Q}{(2\pi)^3}\frac{(G_{\rm{E}}(\textit{Q}^2)-1)}{\textit{Q}^2}e^{-i\textit{Q}\cdot\textit{r}},
\end{align} 
where $V_{C}^{pt}(r)\sim 1/r$ is the Coulomb potential and $V_{C}(r)$ the potential caused by the finite size distribution,
we immediately obtain the energy level shift of the atomic S-states with the corresponding wavefunction $\Psi_{S}$ caused by the finite size effect
\begin{align}
\Delta E = \bra{\Psi_{S}} \delta V \ket{\Psi_{S}}=\frac{2}{3}\pi \alpha |\Psi_{S}(0)|^2 r_{\rm{E}}^2.
\end{align} 
Hereby, a Taylor expansion of the electric form factor 
\begin{align}
G_{\rm{E}}(\textit{Q}^2)&=\int e^{i \textit{Qr}}\rho_{\rm{E}}(\textit{r})   d^3\textit{r}\approx\int \left(1+i\textit{Qr}-\frac{(\textit{Qr})^2}{2}+...\right)\rho_{\rm{E}}(\textit{r}) d^3\textit{r}\\&=1-\frac{1}{6}\textit{Q}^2r_{\rm{E}}^2+\frac{1}{24}\textit{Q}^4r_{\rm{E}}^4+...\approx 1-\frac{1}{6}\textit{Q}^2r_{\rm{E}}^2
\label{Eq5}
\end{align}
has been used.

\subsubsection{Hydrogen $^{1}_{1}$$\rm{H}$}\

\noindent Up to the year 2010, the nuclear charge radius of the proton has been extracted solely from either hydrogen spectroscopy or from electron scattering. Current literature values determined from both methods (hydrogen spectroscopy: $r_{\rm{E}}^{^{1}\rm{H}}=0.8764(89)\,$fm, CODATA-2014 \cite{Moh2016} and electron scattering: $r_{\rm{E}}^{^{1}\rm{H}}=0.8775(50)\,$fm \cite{Moh2016, Sick2014}) agree within their uncertainties. However, they deviate from the value determined from laser spectroscopy of the exotic muonic hydrogen atom $\mu$p ($r_{\rm{E}}^{^{1}\rm{H}}=0.8409(4)\,$fm) \cite{Poh2010}.  Several ideas to explain this discrepancy have been discussed in the meantime \cite{Hill2010, Sick2012, Car2012, Poh2013, Car2015, Ant20162} and gave rise to a re-evaluation of electron scattering data \cite{Sick2014, Ber2014, Lee2015}. One problem could originate from the determination of the Rydberg constant~\cite{Poh2017} from hydrogen spectroscopy and indeed a recent measurement of the Rydberg constant in ordinary hydrogen performed in Garching~\cite{Bey2017} suggested a smaller value for the Rydberg constant and hence a smaller proton radius ($r_{\rm{E}}^{^{1}\rm{H}}=0.8335(95)\,$fm). Although this value perfectly matches the muonic hydrogen measurements, a more recent measurement of the 1S\,-\,3S transition in hydrogen ($r_{\rm{E}}^{^{1}\rm{H}}=0.877(13)\,$fm) \cite{Fle2018} has confirmed the large proton radius. Obviously, more data is needed in order to shed light on the \textit{proton radius puzzle} in the near future.

\subsubsection{Deuterium $^{2}_{1}$$\rm{D}$}\

\noindent
Interestingly, the same discrepancy does also occur for the deuteron, a two-nucleon system (one proton and one neutron). The charge radius extracted from a muonic deuterium measurement, $r_{\rm{E}}^{^{2}\rm{D}}=2.1256(8)\,$fm, \cite{Ant2013} is 5\,$\sigma$ smaller than the CODATA-2014 value of $r_{\rm{E}}^{^{2}\rm{D}}=2.1413(25)\,$fm. The result reinforces the discrepancy that appears for the proton. This becomes clear if one compares the charge radii with the one obtained from the results of the isotope shift measurement of the 1S\,-\,2S transition in hydrogen and deuterium \cite{Hub1998, Par2010}. By adding the mean-square charge radii difference of the deuteron and the proton,$\braket{r^{2}}_{\rm{E}}^{^{2}\rm{D}}-\braket{r^{2}}_{\rm{E}}^{^{1}\rm{H}}=3.82007(65)\,$fm$^2$, to the proton radius from muon spectroscopy one obtains $r_{\rm{E}}^{^{2}\rm{D,\,iso}}=2.1277(2)\,$fm \cite{Jen2011}. This value is in fair agreement with the value from muonic deuterium and hence demonstrates the consistency of the muonic measurements. 

Recent theoretical calculations \cite{Pach2018} of the three-photon exchange corrections to the nuclear structure yielded an updated value for the  difference of the mean-squared deuteron and proton charge radii of $\braket{r^{2}}_{\rm{E}}^{^{2}\rm{D}}-\braket{r^{2}}_{\rm{E}}^{^{1}\rm{H}}=3.82070(31)\,$fm$^2$, from which a more precise value for $r_{\rm{E}}^{^{2}\rm{D,\,iso}}=2.1278(2)\,$fm was obtained.

\subsubsection{Tritium $^{3}_{1}$$\rm{T}$}\

\noindent

Tritium, the heaviest atomic system in the isotopic chain of hydrogen, is composed of a proton and two neutrons and thus represents a three-nucleon system. However, among the lightest elements, its charge radius ($r_{\rm{E}}^{^{3}\rm{T}}=1.755(86)\,$fm) measured by electron scattering reached an accuracy of only 5\,\% \cite{Amr94}. Precision laser spectroscopy of the $^{1}$H\,-\,$^3$T isotope shift of the 1S\,-\,2S transition could be used to improve this value by at least two orders of magnitude \cite{Fri2003}. In combination with the charge radius of its mirror nucleus $^3$He, a distinct improvement of the tritium charge radius could be used for tests of nuclear structure calculations of three-nucleon (3N) forces \cite{Epe2002, Heb2010} and would allow for precision studies of isospin effects \cite{Pia2013, Van2017}.

\begin{figure}[h]
\centering
\subfigure{\includegraphics[width=0.49\textwidth]{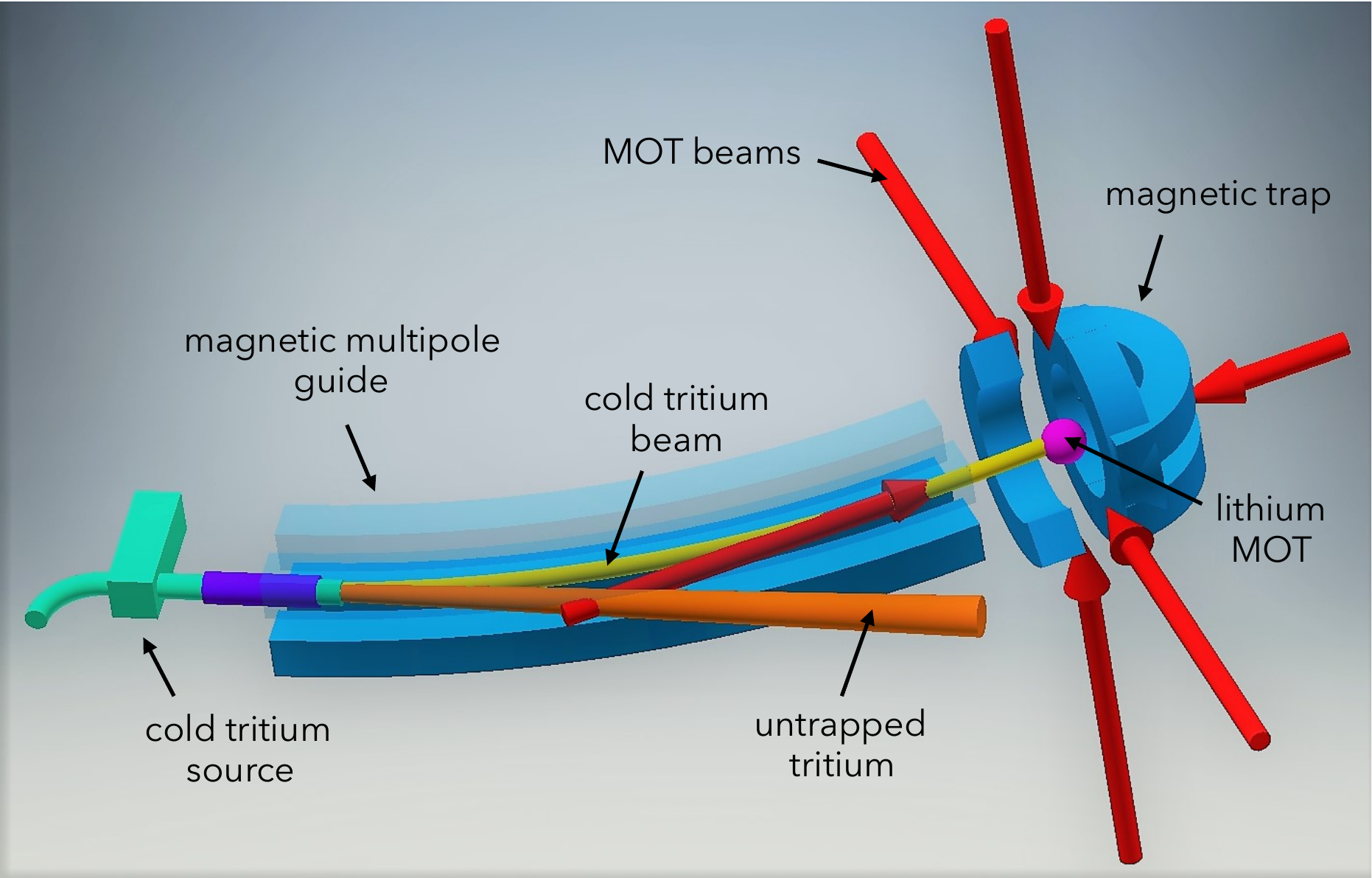}} 
    \subfigure{\includegraphics[width=0.50\textwidth]{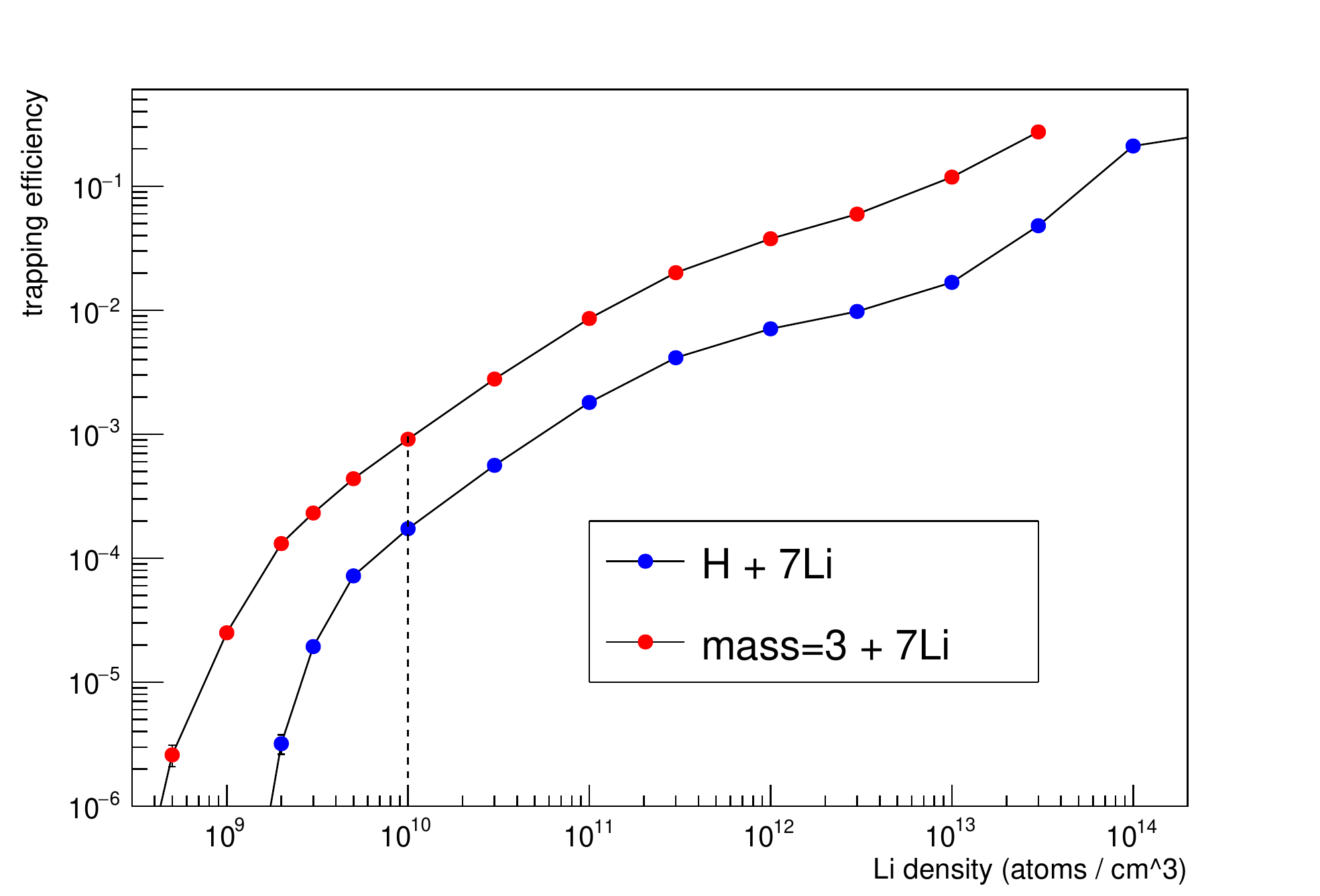}} 
\caption{\label{FigTri} Left: Schematic drawing of the planned apparatus for laser spectroscopy of tritium. Cold atoms escape from a cryogenic nozzle at 4.2\,K. The slow tail of the Maxwell-Boltzmann distribution (yellow) is guided by a magnetic multipole field and injected into a magnetic trap of 0.4\,T. Here, the atoms collide with a cold lithium sample provided by a magneto-optical trap (MOT), which acts as a cold target for buffer gas cooling inside the magnetic trap. Right: Simulated trapping efficiency as a function of the density of the lithium MOT at a given magnetic trap depth of 0.5\,T, using the cross sections of \cite{Cote2000} for H\,-\,Li collisions. Above a density of 10$^{10}$\,atoms per cm$^{3}$ a significant amount of tritium atoms should be trapped. Note that compared to H\,-\,Li scattering the trapping efficiency for T\,-\,Li increases, due to the more favourable mass ratio.}
\end{figure}

In order to provide an improved value for the charge radius of tritium, the Pohl group at the Johannes Gutenberg-University of Mainz aims to perform a high precision laser spectroscopy measurement on an atomic cloud of regular tritium confined in a magnetic potential. The idea exploits the fact that enhanced cooling of trapped hydrogen by a cold lithium target should be feasible, as suggested by calculations \cite{Cote2000}. DeCarvalho and coworkers had suggested this method \cite{DeC2005} to improve cooling of H atoms trapped using superfluid He, but this method is not applicable for T atoms \cite{Wal1990}. This prompted us to consider a dense Li MOT as a cold buffer gas for cooling and trapping of cold T atoms from a 5\,K nozzle \cite{Wal1982}. The fundamental design of our apparatus is presented in Fig.\,\ref{FigTri}. A cryogenic source is used to provide a cold atomic beam. Subsequently, a magnetic multipole guide is used to select the low-velocity part of the beam and to direct the atoms towards a magnetic minimum trap of 0.4\,T. Inside the trap, an ensemble of about 10$^{10}$ lithium atoms serves as a cold target, leading to an immediate energy reduction and trapping of the tritium atoms. The s-wave scattering cross sections for collisions of tritium and lithium are assummed to be similar to the ones for hydrogen and lithium, which have been calculated in \cite{Cote2000} and they appear to be hundred times larger than for H\,-\,H scattering. Based on these values for the scattering cross sections, we have performed theoretical calculations of the stopping process. The results show that an atom density for lithium of about 10$^{10}$ atoms per cm$^{3}$ should be sufficient for the presented approach. As a final step, the lithium atoms are removed from the trap, before a precision measurement on tritium can be performed. With this method, we aim to determine the charge radius of tritium with an 300-fold improved precision.

\subsection{Nuclear Zemach radii}

The nuclear magnetic and electric form factors can be measured directly from electron-proton scattering. On the atomic physics side, however, the magnetic spin-spin interaction between the nucleus and the orbiting lepton gives rise to the hyperfine splitting (HFS). As first noted by Zemach \cite{Zem1956}, the relevant nuclear structure parameter deduced from the HFS is the so-called \textit{Zemach radius}. It is defined as the convolution of the electric and magnetic form factors, $G_{\rm{E}}$ and $G_{\rm{M}}$ respectively,
\begin{align}
r_{\rm{Z}}=-\frac{4}{\pi}\int_{0}^{\infty}\frac{dQ}{Q^2}\left( \frac{\mu_{\rm{N}}}{\mu_{\rm{X}}}G_{\rm{E}}(Q^2)G_{\rm{M}}(Q^2) -1\right)
\label{EqZ}
\end{align}
or, non-relativistically, a convolution of the charge $\rho_{\rm{E}}(\textit{r})$ and magnetization $\rho_{\rm{M}}(\textit{r})$ density 
\begin{align}
r_{\rm{Z}}=\int |\textit{r}| \rm{d}^3\textit{r}\int \rho_{\rm{E}}(\textit{r}-\textit{r}')\rho_{\rm{M}}(\textit{r}') \rm{d}^3\textit{r}'.
\end{align}
In Eq.\,\ref{EqZ} we have used the nuclear magneton $\mu_{\rm{N}}$ and the magnetic dipole moment of the nucleus of investigation $\mu_{\rm{X}}$.

Experimentally, the proton's Zemach radius can be determined through a measurement of the form factors or via the hyperfine splitting $\Delta E_{\rm{HFS}}$ (1S-HFS) in either hydrogen or $\mu$p. The ground-state HFS of muonic hydrogen can be summarized as \cite{Tom200018}
\begin{align}
\Delta E^\mathrm{th}_\mathrm{HFS} = 183.978(16) - 1.287\,r_Z ~\mathrm{[meV]},
\label{rZ_Epol_Eavm}
\end{align}

where $r_{Z}$ is expressed in fermi. This value is in good agreement with the one based on Ref.\,\cite{Pes2017}~\footnote{Eq.\,(\ref{rZ_Epol_Eavm2}) was obtained from the values given in \cite{Pes2017} and assuming a proton Zeemach radiuis of $r_{\rm{Z}}^{^{1}\rm{H}}=1.045(4)\,$fm \cite{Ber2011}. Hereby we conservatively assumed that the uncertainty of the polarizability contribution (included in the first term in Eqs.\,(\ref{rZ_Epol_Eavm}) and (\ref{rZ_Epol_Eavm2})) is equal to the uncertainty of the total TPE of \cite{Pes2017}.}, 
\begin{align}
\Delta E^\mathrm{th}_\mathrm{HFS} = 183.967(21) - 1.287\,r_Z ~\mathrm{[meV]}.
\label{rZ_Epol_Eavm2}
\end{align}

The uncertainty of the first terms is arising mainly from the uncertainty of the polarizability contribution \cite{Faustov2002, Faustov2005, Mart2005, Carlson2008, Carlson2011, Hag2015, Tom2017, Tom2018, HagelsteinMiskimen2017}. The contributions from QED \cite{Eid2007}, weak \cite{Eides2012} and hadronic vacuum polarization \cite{Hag2017, Pes2017} are also included in the first terms. Meson exchange contributions \cite{Dor2017, Dor20172, Hag2018, Len2018} have been discussed. These contributions are however already included in the two photon exchange contributions computed dispersively \cite{Pascalutsa200018, Pineda200018, Vanderhaeghen200018}.


A measurement of the Zemach radius on the $1\,\%$ level or better will influence two aspects of fundamental physics:  nuclear structure theory of the simplest nuclei as well as tests of bound-state QED. The latter results from the fact that the theoretical prediction of the 1S-HFS in e.\,g.~hydrogen (21\,cm line) is currently limited by the proton structure \cite{Eid2007}, \begin{align}
\nu_{\rm{theo}}=1\,420\,403.1(6)_{\rm{proton\,size}}(4)_{\rm{pol}}\,\rm{kHz},
\end{align}
whereas the experimental value has been determined to 12 digits already in the 1970's \cite{Ess1971}, 
\begin{align}
\nu_{\rm{exp}}=1\,420\,405.751\,766\,7(10)\,\rm{kHz}.
\end{align}

\begin{figure}[h]
\centering
\includegraphics[width=0.7\textwidth]{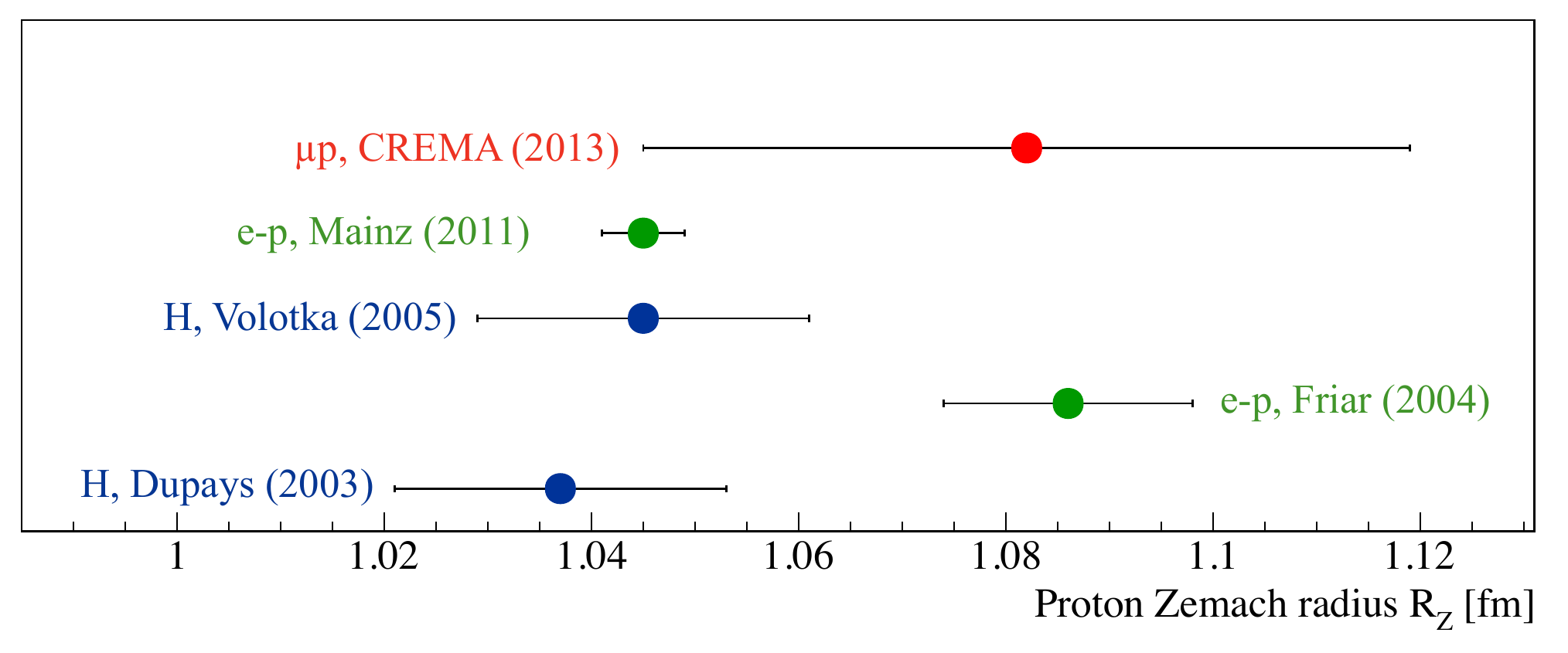}
\caption{\label{FigZemach} Zemach radius of the proton extracted from hydrogen spectroscopy (blue) \cite{Dup2003,Vol2005}, electron scattering data (green) \cite{Fri2004, Ber2011} and from laser spectroscopy of the 2S\,-\,2P transition in muonic hydrogen (red) \cite{Ant2013, Antog2013}.}
\end{figure}

Recent determinations of the Zemach radius from atomic hydrogen yield values of $r_{\rm{Z}}^{^{1}\rm{H}}=1.037(16)\,$fm \cite{Dup2003} and $r_{\rm{Z}}^{^{1}\rm{H}}=1.047(16)\,$fm \cite{Vol2005}, respectively. The value determined from electron scattering experiments was found to be $r_{\rm{Z}}^{^{1}\rm{H}}=1.086(12)\,$fm \cite{Fri2004}.  Later this value was re-measured ($r_{\rm{Z}}^{^{1}\rm{H}}=1.045(4)\,$fm \cite{Ber2011}), showing a good consistency with the hydrogen data (see Fig.\,\ref{FigZemach}). In 2013, a first value of the Zemach radius ($r_{\rm{Z}}^{^{1}\rm{H}}=1.082(37)\,$fm) was individually extracted from laser spectroscopy of the 2S\,-\,2P transition in muonic hydrogen \cite{Ant2013, Antog2013}. In order to improve the value for the Zemach radius, a direct measurement of $\Delta E^{\rm{HFS}}_{1S}$ is foreseen by different groups \cite{Ada2012, Ma2016, Ada2016, Ant2016} using muonic hydrogen spectroscopy.

\subsection{CREMA collaboration - experiment proposal}
\label{SecZ1}

\begin{figure}[h]
\centering
\includegraphics[width=0.95\textwidth]{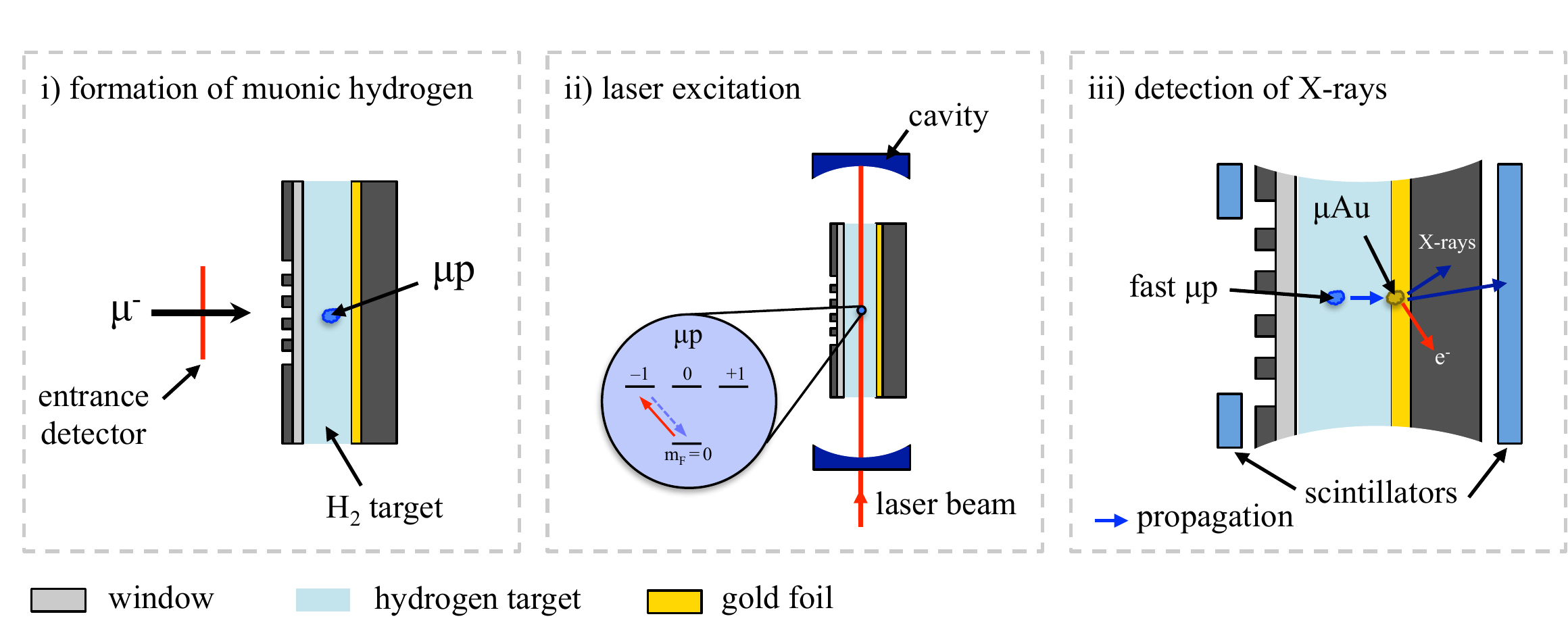}
\caption{\label{Fig4} Sketch of the experimental scheme for the $1S$-HFS spectroscopy in muonic hydrogen.}
\end{figure}

As already described in \cite{Ant2016}, we propose to measure the proton Zemach radius at the Paul Scherrer Institute (PSI), Switzerland, on a 0.25$\,\%$ level, assuming that the polarizability contribution $\delta^{\rm{pol}}$ can be improved to a 5$\,\%$ relative accuracy. The underlying measurement scheme can be divided into three steps (see Fig.\,\ref{Fig4}): i) formation and de-excitation of muonic hydrogen in the $F=0$ ground-state, ii) laser excitation which results in a population transfer to the $F=1$ state and iii) detection of fast de-excited muonic hydrogen by muon transfer from $\mu$p to high-Z muonic atoms in the target walls. In step i), muons with an initial momentum of about $10\,$-$\,12\,$MeV/c are degraded in a $30\,\mu$m thick plastic scintillator acting as an entrance detector. Subsequently, after passing a thin titanium window, about $20\,\%$ of the incoming muons are stopped inside a 2\,mm long hydrogen gas target \cite{Ant2018}. 

After stopping the muons inside the gas target, highly excited muonic hydrogen is formed followed by a prompt muonic cascade to the 1S state. Rapid collisional quenching of the $F=1$ state \cite{Coh1991} leads to a population of the ground-state ($F=0$) in muonic hydrogen in about 200\,ns. About 1.5$\,\mu$s after the muons have entered the target region, a high-intensity laser pulse at $6.7\,\mu$m ($\sim 0.184\,$eV) with a pulse energy of about $3\,$mJ is used to drive the ground-state transition (step ii)).  On resonance, the muon is transferred into the upper hyperfine state ($F=1$). Fast collisions with the surrounding hydrogen gas molecules causes again a de-excitation into the ground state. In this process, about $0.1\,$eV of the HFS transition energy of $0.185\,$eV is converted into kinetic energy of the muonic atom \cite{Coh1991}. 

In step iii), these fast muonic hydrogen atoms propagate towards the target walls \cite{Abb1997}, that are coated with high-Z material. At the wall the muon is transferred to the high-Z atom (e.\,g.~Au) forming a high-Z muonic atom in an excited state. This transfer process is monitored via scintillators by the detection of X-rays emitted during the de-excitation of $\mu$Z. By repeating steps i)-iii) for different laser frequencies a resonance curve is obtained. The number of background events caused by the diffusion of non-excited muonic atoms to the target walls is minimized by cooling the target to about 30\,K.

\section{Helium}

\subsection{Nuclear charge radii}

It's been now more than a decade ago since the first measurements of the nuclear charge radii of the neutron-rich isotopes $^{6}$He \cite{Wan2004} and $^{8}$He \cite{Mue2007} have led to a profound understanding of the nuclear structure in the isotopic chain of helium \cite{Bla2013}. Based on new input from a direct mass measurement of these halo nuclei, the charge radii published in \cite{Wan2004, Mue2007} have been re-evaluated in 2012 \cite{Bro2012}, yielding $r_{\rm{E}}^{^{6}\rm{He}}=2.060(8)\,$fm and $r_{\rm{E}}^{^{8}\rm{He}}=1.959(16)\,$fm, respectively. Here, the absolute charge radius of the stable $^{4}$He isotope is used as a reference ($r_{\rm{E}}^{^{4}\rm{He}}=1.681(4)\,$fm) \cite{Sick2008, Sic2014}. It is extracted from electron scattering data just as the charge radius of the lighter stable helium isotope $^{3}$He ($r_{\rm{E}}^{^{3}\rm{He}}=1.973(14)\,$fm) \cite{Sic2014}.

In the near future new absolute values for the charge radii of the stable helium isotopes will be available from our muonic helium spectroscopy \cite{Anto2010}. The underlying theoretical framework needed to extract the charge radii from the 2S\,-\,2P Lamb shift in helium is summarized in \cite{Fra2017, Die2016}. The charge radii will be determined with a precision improved by an order of magnitude, compared to the current literature values. The results will give new insights into the \textit{proton radius puzzle} and can be used as benchmarks for advanced nuclear structure calculation \cite{Nev2016}, especially of 3N forces. Further it can be used as an independent value for the mean-square charge radii difference of both stable isotopes. This value will help to resolve the discrepancy between different measurements in electronic helium \cite{Shin1995, Roo2011, Pas2012, Pat2016, Pat2017, Ren2018}.

\subsection{Nuclear (magnetic) Zemach radius of $^{3}$$\rm{He}$}

The Zemach radius of $^{3}$He has been deduced from elastic electron-helium scattering. Its value is $r_{\rm{Z}}^{^{3}\rm{He}}=2.528(16)\,$fm \cite{Sic2014}. Based on this value, a prediction of the 1S-HFS ground-state transition energy can be made.  However, there are no theoretical predictions for the HFS polarizability in $\mu^3$He$^{+}$ (see e.g.\,\cite{Dup2003, Mar2005}) so far, leading to only a rough estimate of the transition wavelength of about 930\,nm.

\subsection{CREMA collaboration - experiment proposal}
\begin{figure}[h]
\centering
\includegraphics[width=0.95\textwidth]{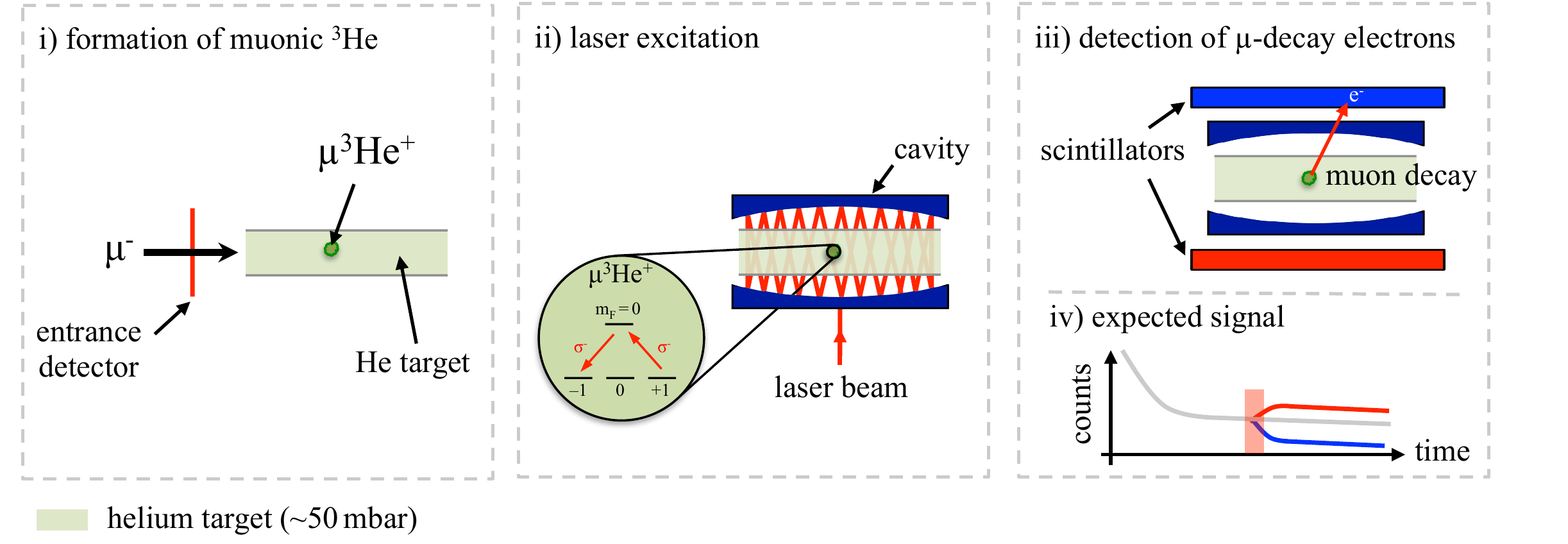}
\caption{\label{Fig5} Sketch of a possible experimental scheme for the 1S-HFS spectroscopy in muonic helium. In iv), the expected  time spectrum for the muon decay asymmetry is shown.}
\end{figure}

A way to determine the Zemach radius of the helion with high precision could be by measuring the ground-state HFS in the muonic ion ($\mu^3$He$^{+}$) \cite{Ant2016}. However, the method described in Sec.\,\ref{SecZ1} cannot be applied to $\mu^3$He$^{+}$, because collisional de-excitation is negligible for this system \cite{Ack1998}. In contrast, our approach to measure the 1S-HFS in $\mu^3$He$^{+}$ relies on the muon decay asymmetry method also applied to $\mu$H by the J-PARC collaboration \cite{Str2018}. In brief, a muon beam is stopped inside a $^3$He gas target, whereby $\mu^3$He$^{+}$ is formed (step i) in Fig.\,\ref{Fig5}). The target section used here is quite similar to the one of the CREMA Lamb shift experiment \cite{Poh2010}. Once the muons are decayed into the ground state, a laser pulse at $930\,$nm is used to drive the ground-state hyperfine transition, inducing an imbalance of the populations of the Zeeman sub-levels.
In step iii), the emission of the electron decay occurs anti-parallel to the spin orientation of the muon due to the parity violation in the muon decay. On resonance, this leads to an asymmetric decay signal (see Fig.\,\ref{Fig5} iv)) in a pair of scintillator detectors, which are mounted around the target region. A resonance curve is obtained by plotting the asymmetry as a function of the laser frequency. 

\section{Lithium}

\subsection{Nuclear charge radii}

Lithium has two naturally occurring isotopes, $^{6}$Li and $^{7}$Li, and three short-lived isotopes on the neutron-rich side of the nuclear chart. Due to its rather simple level structure ($1s{^2}2s$) it represents an ideal candidate for many  high-precision laser spectroscopy experiments \cite{Yan1997, Das2007} including quantum interference studies \cite{Bro2013}. The neutron-rich isotopes $^{8,\,9,\,11}$Li have been synthesized at e.\,g.~TRIUMF, Vancouver or GSI, Darmstadt, and investigated using two-photon laser spectroscopy in combination with RIMS (resonant ionization mass spectrometry) to yield precise values for the isotope shifts \cite{San2006, Ewa2004, Pohl2001H}. As reviewed in \cite{Noe2011}, the current literature values for the absolute charge radii of the lithium isotopes $^{7,\,8,\,9,\,11}$Li (see Fig.\,\ref{Fig0}) were determined relative to the anchor nucleus $^{6}$Li ($r_{\rm{E}}^{^6\rm{Li}}=2.589(39)\,$fm), which has been been obtained from elastic electron scattering data. The uncertainty of the reference isotope represents the dominating contribution to the uncertainty of the charge radii of all other lithium isotopes. Therefore, a new absolute value for the charge radius is required in order to get new insights into the internal nuclear structure of lithium, which for e.\,g.\,the halo nuclei $^{11}$Li is not fully understood \cite{Noe2011}. Laser spectroscopy of the Lamb shift in muonic lithium ($\mu^{6}$Li$^{2+}$ or $\mu^{7}$Li$^{2+}$) could provide a new reference value with an improved accuracy of about an order of magnitude only limited by the calculated value of the nuclear polarizability. The theoretical background and the concept of the planned experiment is described in the following.

\subsection{Theory}

\begin{figure}[h]
\centering
\includegraphics[width=0.8\textwidth]{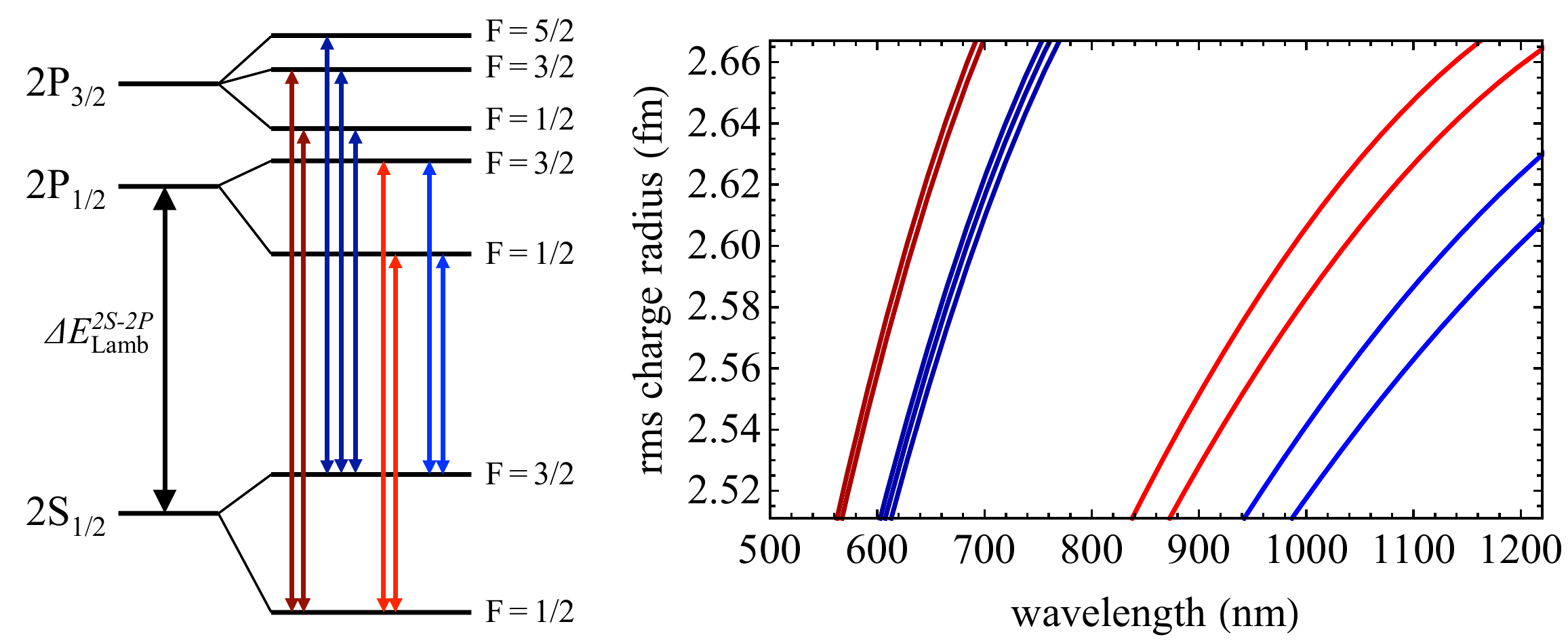}
\caption{\label{Fig1} Left: Level scheme of the $2S\rightarrow2P$ transitions in muonic lithium. Right: Corresponding transition wavelengths as a function of the rms nuclear charge radius of $^{6}$Li. The y-axis spans a $\pm\,2\,\sigma$ interval of $r^{^{6}\rm{Li}}_{\rm{E}}=2.589(39)\,$fm. The calculations are based on the values presented in \cite{Swa1986}. The transitions are color coded according to their 2S$_{1/2}$ F-state.}
\end{figure}

First accurate calculations of the Lamb shift and fine-structure splitting of muonic atoms with $Z \geq 3$ were performed already in the eigthies by Drake et al.~\cite{Dra1985, Swa1986}. These calculations were recently improved by Krutov et al. \cite{Kru2016}. As an example, Fig.\,\ref{Fig1} shows the level scheme of the 2S\,-\,2P transitions in $\mu^{6}$Li$^{2+}$. Due to the finite size effect, the transition wavelength of the Lamb shift exhibits a strong dependence on the size of the nucleus, as shown in the right part of Fig.\,\ref{Fig1}. Transition wavelengths can be found in the yellow ($\sim 600\,$nm) and infrared regime ($\sim 1000\,$nm). The linewidth $\Gamma^{2P}_{\nu}$ of these transitions is to lowest order given by  \cite{Swa1986}\footnote{Please note the corrected pre-factor in Eq.\,\ref{Gamma123}.}
\begin{align}
\label{Gamma123}
\Gamma^{2P}_{\nu}=4.124\times10^{-4}\frac{m_{\rm{red}}}{m_{e}}Z^4 \:\text{[meV]},
\end{align}
where $m_{\rm{red}}/m{_e}$ is the ratio of the reduced mass of the muonic atom and the electron mass. For $\mu^{6}$Li$^{2+}$ we find $m_{\rm{red}}/m{_e}=202.940$ and thus $\Gamma^{2P}_{\nu}=1.64$\,THz, which corresponds to a linewidth of about 1.4\,nm (5.5\,nm) at 600\,nm (1000\,nm).

The lifetime of the (metastable) 2S state in $\mu^{6}$Li$^{2+}$ is 830 ns, determined by the 2-photon decay rate of $1.65\cdot10^3\times Z^6\,$ s$^{-1}$ \cite{Dra1985}. This is long enough to drive the Lamb shift transition\cite{Antog2005, Antog2009}. For comparison, the laser system used for the muonic hydrogen and helium experiments produced laser pulses of about 300\,ns after a muon trigger. In addition, based on the 2S\,-\,2P transition probabilities, we have calculated the saturation fluence $F_{\rm{sat}}$ that is required to drive the individual transitions for all light muonic atoms. Since $F_{\rm{sat}}$ scales with $m_{\rm{red}}^3Z^6$, the saturation fluence of the strongest transition in $\mu^{6}$Li$^{2+}$ ($2S_{1/2}^{F=3/2}\rightarrow 2P_{3/2}^{F=5/2}$) is found to be 17.4\,J/cm$^2$. This can easily be achieved using the existing Yb:YAG pump laser system \cite{Antog2009}. To complete the picture, the properties of the strongest transitions of $\mu^{7}$Li$^{2+}$ and $\mu^{9}$Be$^{2+}$ are listed in Tab.\,\ref{Tab2} accordingly.

 \begin{center}
\begin{table}[h]
\caption{\label{Tab2} $2S\rightarrow 2P$ transition properties of light muonic atoms. Only the strongest transitions are presented. The linewidth $\Gamma^{2P}_{\nu}$, the reduced mass $m_{\rm{red}}$, the nuclear spin $I$, the laser saturation fluence $F_{\rm{sat}}$, the lifetime of the $2S$ state $\tau_{2S}$ and the transition wavelength $\lambda$ are given. $^{\dagger}$ Limited by collisional quenching at 1\,mbar gas pressure \cite{Pohl2006}. $^{\ddagger}$ Only slightly affected by quenching at ion gas pressure \cite{Arb1984}.}
\centering
\begin{tabular}{@{}*{9}{l c c c c c c c  c c c  }}
\br
isotope & transition & $\Gamma^{2P}_{\nu}$\,(GHz) & $m_{\rm{red}}/m{_e}$ & $I$ & $F_{\rm{sat}}$\,(J/cm$^2$) & $\tau_{2S}$\,(ns) & $\lambda$\,(nm) \\
\mr
$\mu^{1}$H & $2S_{1/2}^{F=1}\rightarrow 2P_{3/2}^{F=2}$ & 18.5 & 185.8 & 1/2 & 0.0165 & 1000$^{\dagger}$ & $\sim6010$ \\
$\mu^{2}$D & $2S_{1/2}^{F=2}\rightarrow 2P_{3/2}^{F=3}$ & 19.5 & 195.7 & 1 & 0.0165 & 1000$^{\dagger}$ & $\sim5900$  \\
\mr
\mr
$\mu^{3}$He$^{+}$ & $2S_{1/2}^{F=0}\rightarrow 2P_{3/2}^{F=1}$ & 318 & 199.2 & 1/2 & 1.11 & 1700$^{\ddagger}$ & $\sim 860$ \\
$\mu^{4}$He$^{+}$ & $2S_{1/2}^{F=1}\rightarrow 2P_{3/2}^{F=2}$ & 321 & 201.1 & 0 & 1.11 & 1700$^{\ddagger}$ & $\sim 810$\\
\mr
\mr
$\mu^{6}$Li$^{2+}$ & $2S_{1/2}^{F=3/2}\rightarrow 2P_{3/2}^{F=5/2}$ & 1639 & 202.9 & 1 & 17.4 & 830 & 600\,-\,700 \\
$\mu^{7}$Li$^{2+}$ & $2S_{1/2}^{F=2}\rightarrow 2P_{3/2}^{F=3}$ & 1644 & 203.5 & 3/2 & 18.8 & 830 & 600\,-\,650 \\
\mr
\mr
$\mu^{9}$Be$^{3+}$ & $2S_{1/2}^{F=2}\rightarrow 2P_{3/2}^{F=3}$ & 5213 & 204.2 & 3/2 & 106.5 & 150 & 830\,-\,1030\\
\br
\end{tabular}
\end{table}
\end{center}

\subsection{Concept of an experimental apparatus}

\begin{figure}[h]
\centering
\includegraphics[width=0.95\textwidth]{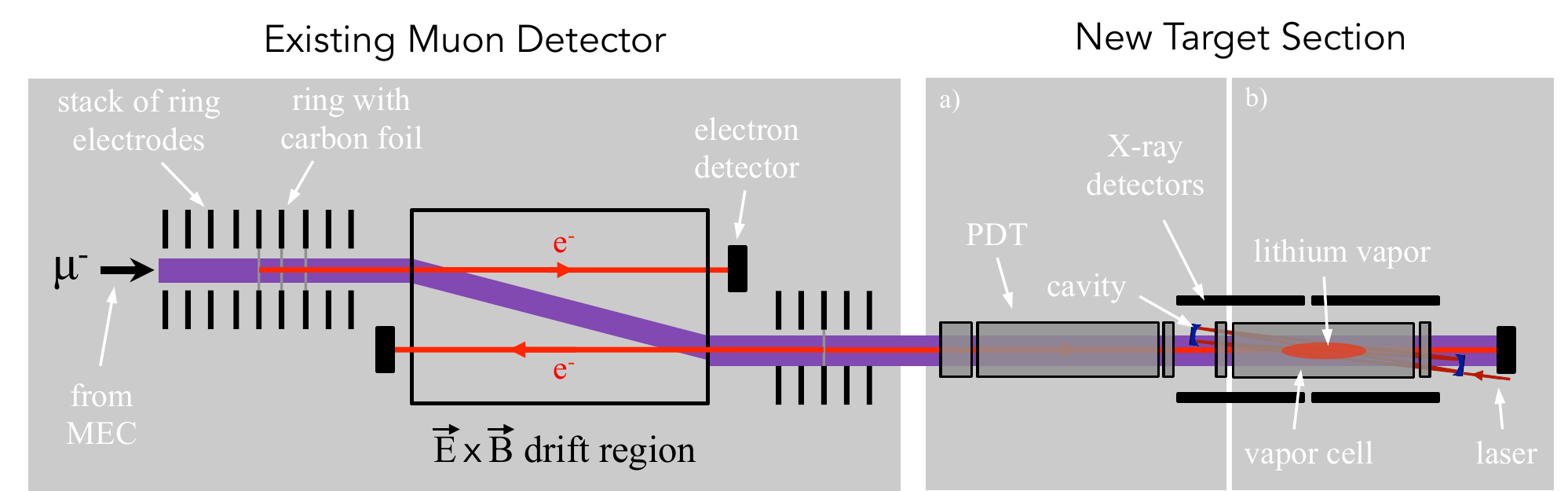}
\caption{\label{Fig6} Schematic drawing of the proposed lithium apparatus inside the 5\,T magnet of the low-energy muon beamline (not to scale). PDT: pulsed drift tube, MEC: muon extraction channel.}
\end{figure}

One of the main challenges for laser spectroscopy experiments with muonic systems above $Z \geq 3$ is the preparation of a dense, gaseous target from a solid. Such a target could be realized, at least for lithium, by the use of a compact hot vapor cell \cite{Vid1969, Vid1971, Bac1974}, which is embedded in the 5\,T solenoid of the low-energy muon beamline \cite{Sim1992, DeC1997}. A sketch of the proposed target section including the non-destructive muon detection system is depicted in Fig.\,\ref{Fig6}. Here, only a brief description of the muon detector is given. Inside a first stack of ring electrodes muons with a kinetic energy of 20\,keV\,-\,40\,keV are forced to traverse several layers of ultra-thin carbon foils (few $\mu$g/cm$^2$), acting as both a moderator and as an electron source. The trajectory of electrons that are kicked out of the carbon foils by the muon are separated inside a $\vec{E}\,\times \vec{B}$ velocity filter and are subsequently detected outside the drift region using photo-multipliers. By repeating this step in a second stack of ring electrodes a coincidence signal is generated. The non-destructive muon detector thus acts as an arrival detector for muons (for a similar device for ions see e.\,g.~\cite{Sch2015}) and is used to trigger the spectroscopy laser. All this constitutes the existing CREMA setup used very successfully in the muH, muD and muHe measurements.

After detection, the muons leave the muon detector with an energy of $\sim2\,$keV and enter the target section. Here, they are further decelerated by either fast switching of a pulsed drift tube (PDT), by applying a negative bias voltage to the hot vapor cell, or both. By this means, an ultra-slow muon beam with a kinetic energy of far below 100\,eV can be produced that is finally injected into a dense cloud of lithium atoms. First estimates have shown that a vapor pressure of 1$\,\mu$bar ($n^{Li}\sim7\cdot10^{12}$\,atoms/cm$^{3}$) should be sufficient to produce a reasonable amount of muonic lithium ions ($\mu^{6}$Li$^{2+}$) inside the gas target (for comparison, a target pressure of $\sim$2\,mbar was used for helium and we have previously stopped muons in a hydrogen gas pressure as low as 0.063\,mbar \cite{Pohl2001t}). This corresponds to a temperature of the heat pipe of about 300\,$^{\circ}$C to 500\,$^{\circ}$C. To protect the surroundings from damages through these high temperatures, the installation of a dedicated cooling system combined with heat shields is foreseen.  For detection of the muonic X-rays a stack of CdTe detectors \cite{amptek} promise to be a good choice, reaching similar or better performance as the APDs used before \cite{Fern2003, Lud2005, Fern2007}.
They provide a detection efficiency of up to 100$\,\%$, with excellent energy and time resolution. In addition, the inner part of the target section needs to be transparent at $E^{Li}_{K_{\alpha}}=18.7\,$eV. 
In contrast to the Lamb shift experiments, the cavity for the spectroscopy laser will be installed slightly tilted from the beam axis to provide a good overlap with the target ions (see Fig.\,\ref{Fig6}).  

\section{Beryllium}

\subsection{Nuclear charge radii}

The isotopic chain of beryllium contains the proton-rich isotope $^7$Be, the stable or reference isotope $^9$Be, and the neutron-rich isotopes $^{10,\,11,\,12}$Be.  Their nuclear charge radii have been measured to high precision in the last years by frequency-comb assisted laser spectroscopy \cite{Kri2012, Kri2018}. The measurements were motivated by the fact that the singly charged beryllium ion is ideally suited for tests of many-body bound-state QED calculations in three-electron systems and nuclear structure calculations. This has led to e.\,g.~sophisticated cluster models of the proton and neutron distribution inside the beryllium nuclei, with major interest in the two halo nuclei $^{11,\,12}$Be. A detailed overview of the nuclear structure of beryllium is provided by Krieger and co-workers \cite{Kri2018}. 

The current literature value for the charge radius of the anchor nucleus $^9$Be ($r_{\rm{E}}^{^9\rm{Be}}=2.519(12)\,$fm) has been determined in 1972 \cite{Jan1972}. Here, we propose to improve the precision of this value by an order of magnitude using muonic beryllium ions ($\mu^9$Be$^{2+}$), while, in principle, laser spectroscopy of muonic beryllium would allow for a 100-fold improvement, the accuracy will be limited by the current level of nuclear polarizability calculations. At the same time, a new reference value would lead to a direct improvement of the precision of the charge radii in the whole isotopic sequence by a factor of $\sim2$.

\subsection{Experimental approach}

\begin{figure}[h]
\centering
\subfigure{\includegraphics[width=0.32\textwidth]{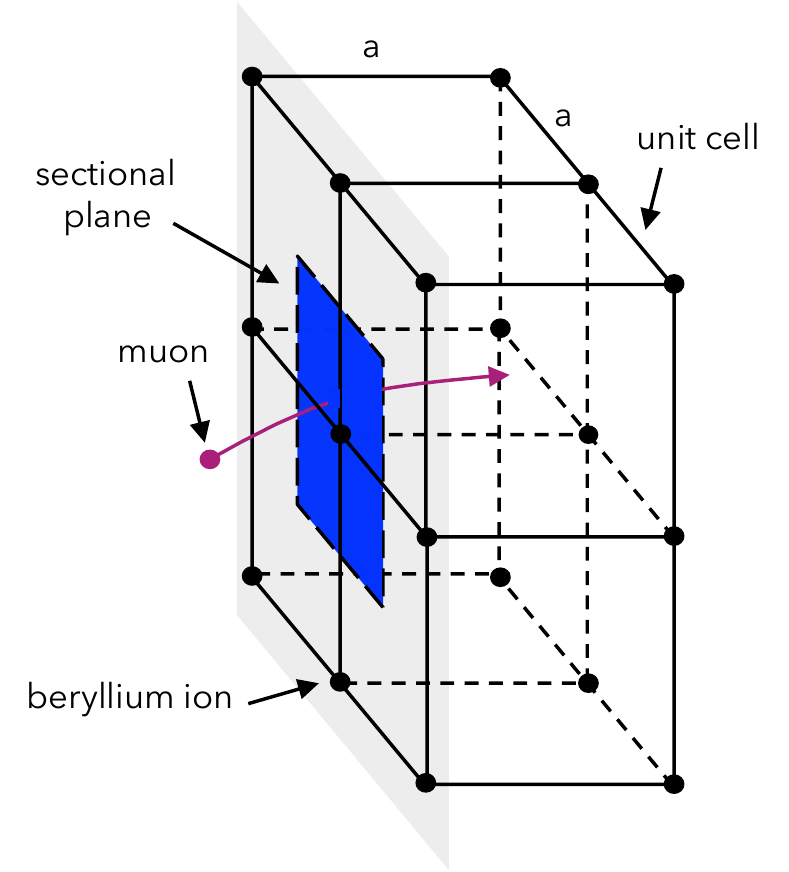}} 
    \subfigure{\includegraphics[width=0.67\textwidth]{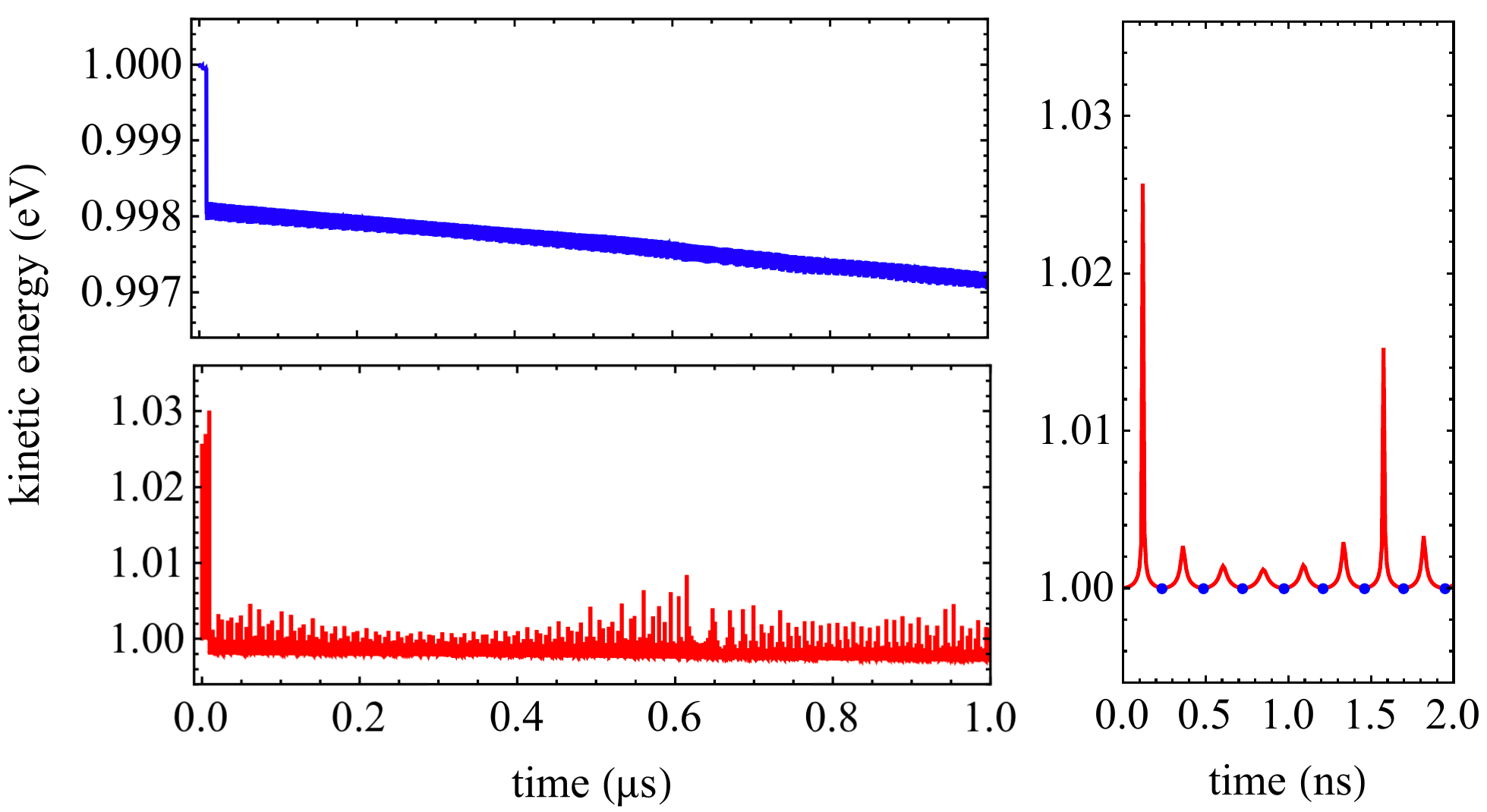}} 
\caption{Left: Simplified picture of the scattering process and involved quantities. Right: Kinetic energy of a muon while passing through a beryllium ion crystal. The timescale of the stopping process is chosen to be half of the muon's lifetime. At a kinetic energy of 1\,eV this corresponds to a path length of about 4\,cm and 3\,000 collisions. The blue data points show the kinetic energy after passing one unit cell, whereas the red data points show the kinetic energy of the muon after each time step.}
\label{Fig7}
\end{figure}

The concept of a hot vapor cell cannot be applied for beryllium. Therefore, we investigate the use of a cold beryllium ion crystal confined in a variation of a Penning(-Malmberg) trap as a dense target. In the following we made a simple estimate of the expected muon capture rate and event rate achievable with this approach. Hereby, the length $L_{p}$ and the density $n^{\rm{Be}}$ of the cigar shaped ion crystal are crucial parameters that enter the calculations. We simplified the calculations by considering a muon traversing many times a unit cell of an ion crystal, as depicted in Fig.\,\ref{Fig7}. Assuming a homogeneous flux of muons entering one unit cell of length $a$ of a simple cubic crystal lattice\footnote{Note: In a real ion crystal confined in a Penning trap the geometrical structure of the lattice might be different. However, at this point the assumption is sufficient to estimate the muon capture rate.} (see Fig.\,\ref{Fig7} left), the probability $w_{\mu}$ for the production of a muonic atom is given by
\begin{align}
w_{\mu}(E_{\rm{kin}}^{\mu})=\frac{\sigma_{\mu}(E_{\rm{kin}}^{\mu})}{A_{\rm{sc}}},
\end{align}
where $A_{\rm{sc}}=a \times a$ is the geometrical cross section of the unit cell and $\sigma_{\mu}(E_{\rm{kin}}^{\mu})$ is the muon capture cross section. For e.g.~$a=10\,\mu$m and $\sigma_{\mu}(1\,\rm{eV})\sim 200\,a_{0}^2$ \cite{Coh2004}, we find $w_{\mu}\sim 5\cdot 10^{-9}$, where the muon's kinetic energy is assumed to be constant throughout the scattering process. The muon beamline usually provides 1\,000 muons per second with energies of $\sim 1$\,keV. When assuming a deceleration efficiency from 1\,keV to 1\,eV of $80\,\%$ and a length of the plasma column of 15\,cm, the muon capture rate amounts to $w_{\mu}(1\,\rm{eV})\sim 5\,\cdot10^{-2}\,$ per second. 

The above mentioned assumption of $10\,\mu$m for the length of the unit cell results from the fact that 10$\,\%$ of the Brillouin density (maximum achievable ion density in the crystal),
\begin{align}
n^{\rm{Be}}_{B}=2.96\cdot 10^{8} \left(\frac{B}{1\,\rm{T}} \right)^2 \:[\rm{cm}^{-3}],
\end{align}
was assumed. For comparison, in \cite{Hol2000} a steady-state confinement of up to 10$^{9}$ magnesium ions on a timescale of weeks in a Penning-Malmberg trap was achieved, with densities of up to 20$\,\%$ of $n_{B}$. 
By taking this as a benchmark and by considering that the muons will loose some energy through Coulomb collisions with the beryllium ions, leading to a strong increase of the capture cross section, one could hope for a muon capture rate of about one muon per second. This corresponds to about one laser-induced X-ray event per hour. Here, we have further taken into account the 2S population probability of $3\,\%$ and an X-ray detection efficiency of $60\,\%$. 

To determine the energy loss of a single muon inside a large beryllium ion crystal we have performed preliminary simulations of the stopping process. Our model is different from those in literature (see e.g.~\cite{Bus2006, Hil2014}) and is mainly based on two assumptions: the muon trajectory is repeatedly computed only within one unit cell and interactions with neighboring ions are (so far) neglected. These assumptions are made to reduce the computation time tremendously. 

The right part of Fig.\,\ref{Fig7} shows the energy loss of a muon transversing an ion crystal with an initial energy of 1$\,$eV. Single collisions with a target ion take place on a timescale of 0.25\,ns, indicated by a temporal energy increase. The total energy loss for \textit{soft} collisions within one $\mu$s is less than 0.1$\,\%$ of the muon's kinetic energy and thus can be neglected. Occasionally \textit{hard} collisions (not shown in Fig.\,\ref{Fig7}) occur, leading to an energy reduction of some meV.

Antiprotons, such as the slow $\bar{p}$ beam soon available at CERN's new ELENA (Extra Low ENergy Antiprotons) facility \cite{Mau2014} would be the ideal test of the capture process and may pave the way to novel investigations of antiprotonic atom research using cold ions stored in a Penning trap as a target. Upon capture, the fingerprint in the time spectra of the annihilation of an antiproton could be used to measure the formation rate of e.\,g.~antiprotonic beryllium $\bar{p}^9$Be$^{3+}$. Such results would represent an important benchmark for the proposed studies with muonic beryllium.


\section{Conclusion}

In recent years, laser spectroscopy of muonic atoms has become an indispensable tool for resolving the nuclear structure of light nuclei. New values for the absolute charge radii were obtained by this means, which in case for the proton and deuteron show a large discrepancy compared to the CODATA-2014 world average values based on electronic systems. Thus, in this work we have summarized the current literature values for the nuclear charge radii of the lightest elements ranging from hydrogen to beryllium. In addition, we have proposed new measurements of the Lamb shift in muonic lithium and beryllium in order to provide new independent absolute charge radii of these elements with improved accuracy. Being able to stop muons in a Penning trap to form muonic ions, as proposed for muonic beryllium, would enable a wealth of new measurements with virtually any stable or long-lived isotope, by means of sympathetic cooling of an ion crystal using Be$^{+}$, Ca$^{+}$, Mg$^{+}$ or similar cooling ions.

Confronting these radii with results from precision measurements in electronic atoms and elastic electron or muon scattering measurements will allow new tests of QED and the Standard Model. In the case of agreement, the combination of electronic and muonic measurements will dramatically improve our understanding of nuclear charge (and magnetic Zemach) radii, and nuclear polarizabilities \cite{Antog2013, Fra2017, Die2016, Kra2016}. Further, we have presented our approaches towards a precision measurement of the ground-state hyperfine splitting in muonic hydrogen and helium. 


\section*{Acknowledgments}

We thank O.~Tomalak for summarizing the contributions to the ground-state hyperfine splitting in muonic hydrogen. Further, fruitful discussions with S.~Bacca, N.~Barnea, C.~Carlson, M.~Gorchtein, K.~Pachucki, V.~Pascalutsa and M.~Vanderhaeghen are highly acknowledged.
We thank the support of the Cluster of Excellence PRISMA, the Swiss National Foundation, Projects 200021L$\_$138175 and 200021$\_$165854, of the European Research Council ERC  CoG.~$\#$725039 and StG.~$\#$279765, of the Deutsche Forschungsgemeinschaft DFG GR 3172/9-1, of the program PAI Germaine de Sta\"{e}l no.~07819NH
du minist\`{e}re des affaires \'{e}trang\`{e}res France, the Ecole Normale Sup\'{e}rieure (ENS), UPMC, CNRS,
and the Funda\c{c}\~{a}o para a Ci\^{e}ncia e a Tecnologia (FCT, Portugal) through project PTDC/FIS-NUC/1534/2014).

\end{document}